# Geant4 Modeling of Energy and Charge Deposition in Satellites Solar Cells

Youssef Abouhussien*, Gennady Miloshevsky†

*Virginia Commonwealth University, 401 West Main St, Richmond, VA 23284, abouhussieny@vcu.edu
†Virginia Commonwealth University, 401 West Main St, Richmond, VA 23284, gennady@vcu.edu

**ABSTRACT**

Understanding radiation effects in spacecraft components is critical for predicting long-term performance degradation. In this work, a Geant4 Monte Carlo model is developed to compute charge and energy deposition in satellite solar-cell materials exposed to electrons, protons, and X-ray environments. The model is validated against published experimental and computational benchmarks and shows strong agreement across multiple energy ranges. Power-density deposition profiles are then evaluated for a multi-layer solar-cell structure under blackbody soft X-rays, mono-energetic X-rays, and high-energy electrons. The results show that low-energy X-rays dominate damage in surface layers, while high-energy particles penetrate deeper into semiconductor and substrate layers, posing risk to electronic components. These findings highlight the importance of radiation-specific shielding strategies for space solar-cell design.

**INTRODUCTION**

Modeling of radiation damage in spacecraft and satellite material is very important to scientists who are doing research on materials degradation analysis. Satellites in the Earth's orbit are subjected mainly to high-energy electrons trapped in Earth's magnetic field (the Van Allen belts), high-energy protons from solar flares, and high-energy particles from cosmic rays [1]. There's also a potential hazard from nuclear detonation and electromagnetic pulse attack, which would make satellites in Earth's orbit vulnerable and at high risk of damage [1], [2]. These kinds of radiation affect satellite surface materials in many ways, mainly due to charge and energy deposition, which can cause great damage to the surface material of the spacecraft. Charge deposition can accumulate inside insulators and cause electrostatic discharge, while energy deposition can cause mechanical failure [3]. Other studies has calculated the energy deposition effects from X-rays irradiation in the surface materials of satellites [4], [5].

In this paper, we reported our results from a Geant4 code that calculates the charge and energy deposition in satellite surface materials. We validated our code against multiple experimental and other computational data for electrons, protons, and X-rays. Then we calculated power density distribution in a solar cell for blackbody X-ray spectra at temperature of 1 keV and compared it against both high energy electrons and mono-energetic X-ray beams.

**Geant4 Model and Validation**

The energy deposition and charge deposited by particles in materials can be evaluated using the Monte Carlo (MC) approach. Geant4 provides a complete set of tools using MC method for all areas of radiation transport simulation, geometry tracking, and material response [6], [7]. The low-energy Livermore model is used to describe the electromagnetic physics processes for photons and electrons including photo-absorption, Rayleigh scattering, Compton scattering, gamma conversion, impact ionization, X-ray fluorescence, Auger electrons, Auger cascade, as well as generation of secondary particles [8]. Different production cutoffs has been used for validation (1 μm to 1 nm). Validations against experimental and other computational data are illustrated in Fig. 1 through Fig. 5. Aluminum and ammonium perchlorate are widely used in aerospace industry especially in solid propulsion for rockets [9]. The purpose is to validate the code for high energy electrons for both charge and energy deposition, charge deposition by protons as well as energy deposition by soft X-rays, which deposit most of their energy at the surface of materials [2].

The shapes of the curves from Geant4 simulation and the data from experimental and computational results are found to be in very good agreement for charge distribution of electrons (Fig. 1 and Fig. 2), although there's a slight shift in the peak (Fig. 1), which might be due to experimental uncertainty as the data is relatively old. Electron energies of 300 keV and 1 MeV has been chosen to validate the code in low and high energy ranges for different materials.

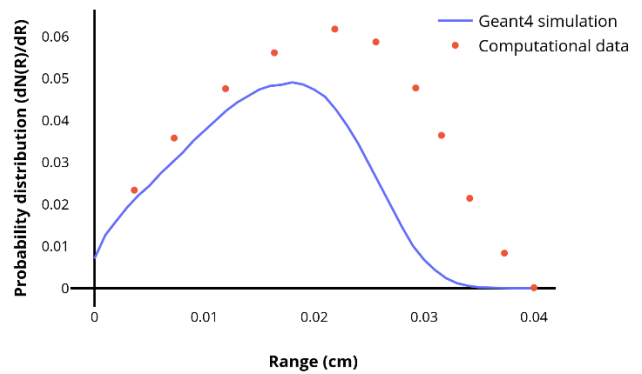

Fig. 1. Probability distribution of 300 keV electrons in aluminum compared against the computational data [9].

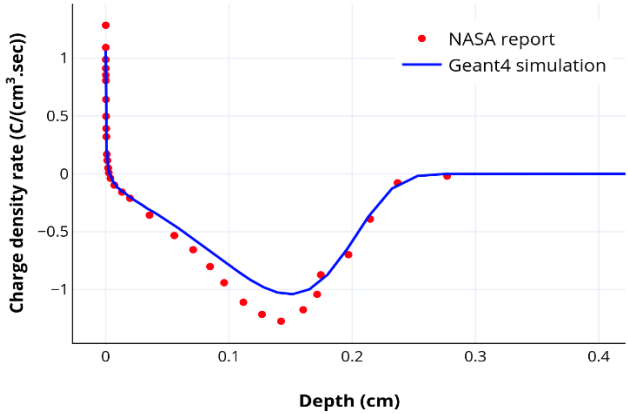

Fig. 2. Charge density rate of 1 MeV electrons in ammonium perchlorate against results from NASA report [9].

Electron charge deposition is low near the surface as secondary generated electrons tend to escape, which can be seen clearly in Fig. 2 as positive charge appears on the surface.

In Figs. 3 and 4, the code has been validated for energy deposition in single and multi-slab system. The energy deposition curves agree well with the experimental data aside from the peaks near the interfaces (Fig. 4), which can be attributed to the uncertainty of the data near the interface boundary, and since the deposition at the interface boundary is considered to belong to the material on the left-side. The units of the y-axis makes the difference very noticeable as the energy is divided by the material's density [10].

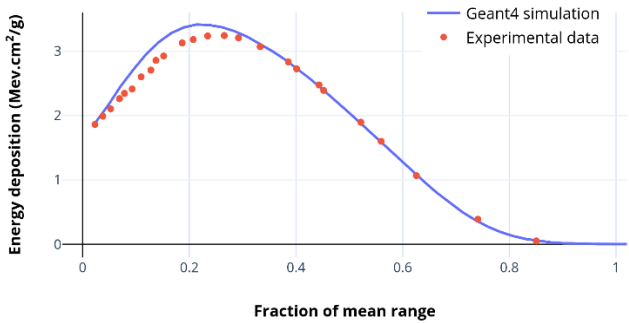

Fig. 3. Energy deposition of 1 MeV electrons in aluminum compared against the experimental data [10]

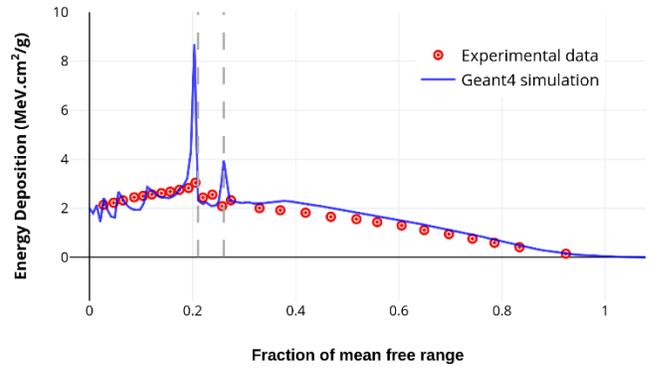

Fig. 4. Energy deposition of 1 MeV electrons in Be-Au-Be slab system [10].

The ability of code to capture positive charge distribution from protons is shown in Fig. 5, the semi-empirical data is for proton's mean range in aluminum, while Geant4 simulation result is for charge distribution. The peak agrees well with the proton range in the material, which should correspond to a positive charge at this mean range. As it's shown in Fig. 5, protons locally deposit all their energy upon first collision within the material and stop there, unlike electrons which deposit their energy in a continuous manner [11].

In Fig. 6, the dose profile agrees well with the computational results from DTRA report using a blackbody X-ray spectrum with temperature of 1 keV (Fig. 7). This energy range is particularly of interest as the soft X-rays damage the surface of the spacecraft and doesn't penetrate deep.

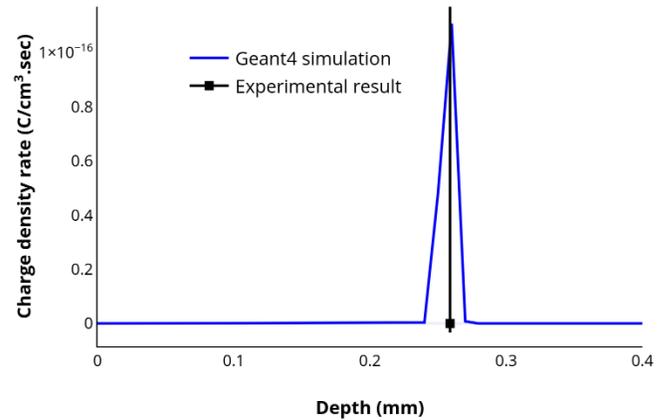

Fig. 5. Charge density rate of 6 MeV protons in aluminum compared against the experimental result for mean proton range [11].

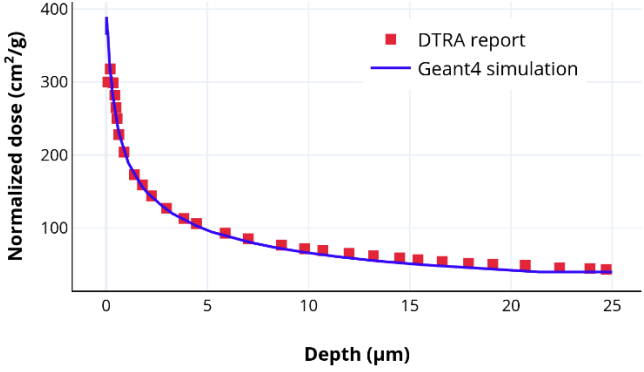

Fig. 6. Normalized dose in poco-graphite for blackbody X-ray spectrum at T = 1 keV compared against the computational results taken from DTRA report [2]

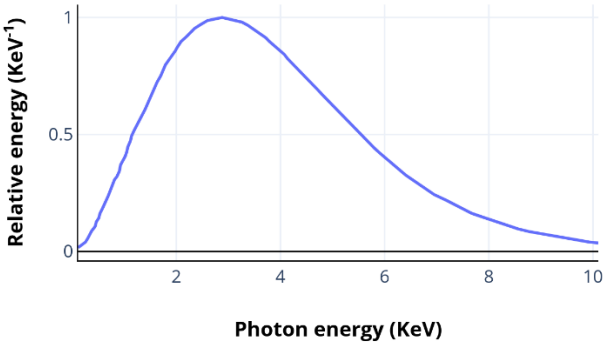

Fig. 7. Black-body spectrum of X-rays at T= 1 keV [2].

In contrast, high energy X-rays penetrate deep into the material where they deposit most of energy and inflict more damage to the electronic components [2].

**RESULTS**

We calculated power density profiles in a multi-layer system of a satellite solar cell with the configuration of layers given in TABLE I. The simulation was performed for different X-ray energies and 1 MeV electrons as shown in Fig. 8. We compared the energy deposition distribution for different radiation environments that the satellite might be vulnerable to. As it's shown in Fig. 8, in the case of X-rays, the outermost 2 layers experience the most power density deposition from a black body spectrum of T = 1 keV (Fig. 6) and then it keeps decreasing as the X-rays don't penetrate much deeper beyond the last Kapton layer. We can see a similar behavior for the mono-energetic 1 keV case, but with higher value. In contrast, for the 1 MeV mono-energetic hard X-rays the power density deposition is much less than the former two cases in the first 2 layers, and starts increasing in the fifth layer. It keeps penetrating deeper beyond the solar-cell layers. In case of 1 MeV electrons, it appears that electrons continuously deposit power density in approximately homogenous manner in the first 6 layers. It continues to deposit deeper beyond the solar cell.

In conclusion, low energy X-rays (1 - 10 keV) induce more damage to the solar-cell than high energy X-rays in terms of power density in the outermost layers. High energy electrons deposit a higher overall power density in the whole solar cell. We can also conclude from the curve in Fig. 8. that the 1 MeV X-rays would deposit most of its energy beyond the solar cell. Therefore, it deposits more energy in the electronic components than low energy X-rays and 1 MeV electrons. It should also be noted that energy deposition isn't the only damaging mechanism in case of electrons. Electrons can deposit its charge inside the solar cell, which increases the risk of local electrostatic discharge in the dielectric materials. Further studies would be needed to investigate the overall effect of high energy particles on the components beyond the solar cell.

TABLE I. Solar cell layers

| Layers (counted from left in Fig. 8) | Thickness, μm |
|---|---|
| 1st Layer: Anti-Reflective (AR) coating, $MgF_2$ | 0.12 |
| 2nd Layer: CMX cover glass | 100 |
| 3rd Layer: Silicone adhesive DC 93-500 | 12 |
| 4th Layer: GaInP | 0.8 |
| 5th Layer: GaAs | 3.6 |
| 6th Layer: Ge | 300 |
| 7th Layer: Silicone adhesive DC 93-500 | 12 |
| 8th Layer: Kapton substrate | 25 |

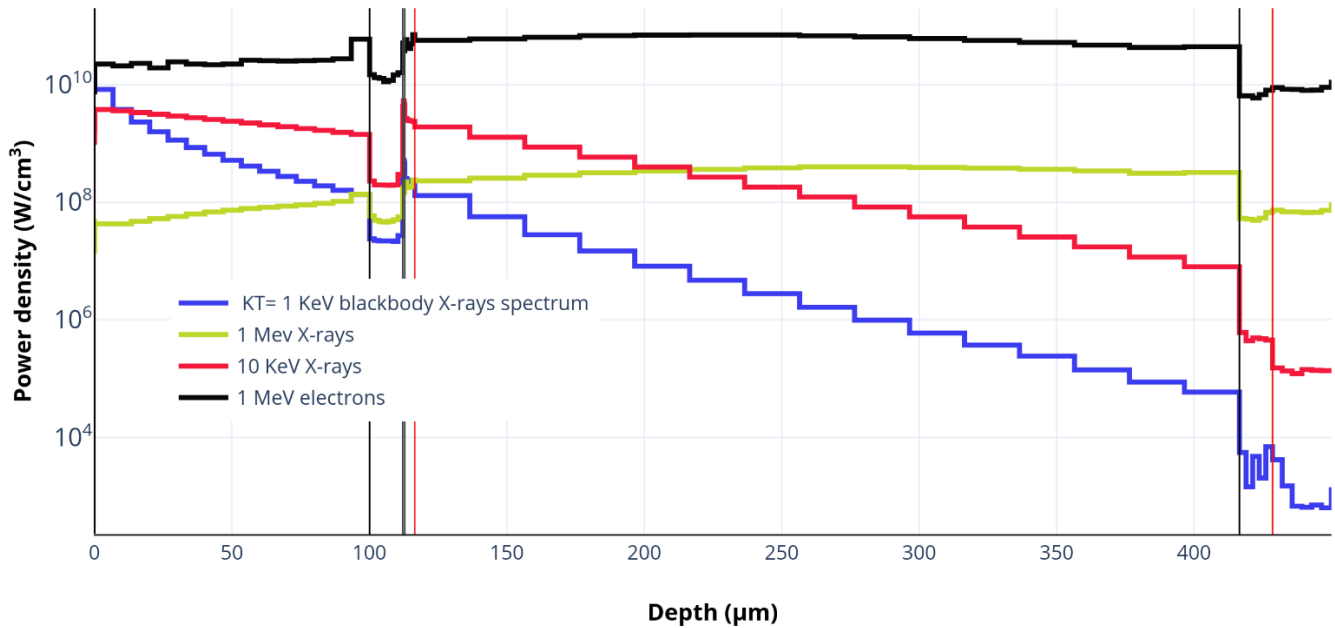

Fig. 8. Power density distribution in a satellite solar cell.

**ACKNOWLEDGMENT**

This work is supported by U.S. Nuclear Regulatory Commission, Grant No. 31310018M0047.